\documentclass{PoS}
\usepackage{enumitem}
\usepackage[below]{placeins}
\title{SiPM and front-end electronics development for Cherenkov light detection}
\ShortTitle{SiPM and front-end electronics development for Cherenkov light detection}
\author{
G.~Ambrosi$^{\;(1)}$, F.~Acerbi$^{\;(2)}$, \speaker{E.~Bissaldi}$^{\;\; \dagger \,(3)}$, A.~Ferri$^{\;(2)}$, F.~Giordano$^{\;(3,4)}$, A.~Gola$^{\;(2)}$, M.~Ionica$^{\;(1)}$, R.~Paoletti$^{\;(5,6)}$, C.~Piemonte$^{\;(2)}$, G.~Paternoster$^{\;(2)}$, D.~Simone$^{\;\;\dagger \;(3)}$, V.~Vagelli$^{\;(1)}$, G.~Zappala$^{\;(3)}$, N.~Zorzi$^{\;(3)}$, for the CTA Consortium$^{\;\;\ddagger}$ \\
$^{(1)}$INFN -- Sezione di Perugia, Perugia, Italy;
$^{(2)}$Fondazione Bruno Kessler (FBK), Trento, Italy;
$^{(3)}$INFN -- Sezione di Bari, Bari, Italy;
$^{(4)}$Dipartimento Interateneo di Fisica, Universit\`a e Politecnico di Bari, Bari, Italy;
$^{(5)}$INFN -- Sezione di  Pisa, Pisa, Italy;
$^{(6)}$ Universit\`a di Siena, Siena, Italy.\\
$\dagger$ {\footnotesize{E-mail:}} {\tt{\footnotesize{Daniela.Simone@ba.infn.it,Elisabetta.Bissaldi@ba.infn.it}}} \\
$\ddagger$ {\footnotesize{Full consortium author list at http://cta-observatory.org}}
}
\abstract{
The Italian Institute of Nuclear Physics (INFN) is involved 
in the development of a demonstrator for a SiPM-based 
camera for the Cherenkov Telescope Array (CTA) experiment, 
with a pixel size of 6$\times$6 mm$^2$. The camera houses 
about two thousands electronics channels and is both 
light and compact. 

In this framework, a R\&D program for the development of 
SiPMs suitable for Cherenkov light detection 
(so called NUV SiPMs) is ongoing. 
Different photosensors have been produced at 
Fondazione Bruno Kessler (FBK), with different micro--cell 
dimensions and fill factors, in different geometrical arrangements. 
At the same time, INFN is developing front--end electronics 
based on the waveform sampling technique optimized for 
the new NUV SiPM.
Measurements on 1$\times$1 mm$^2$, 3$\times$3 mm$^2$, 
and 6$\times$6 mm$^2$ NUV SiPMs coupled to the 
front--end electronics are presented.}

\FullConference{The 34th International Cosmic Ray Conference,\\
		30 July - 6 August, 2015\\
		The Hague, The Netherlands}
\begin{document}

\section{The prototype}
\subsection{The Detectors - Silicon Photomultipliers}

In the following analysis, we present the performances of the Near Ultra--Violet Silicon Photomultipliers (NUV SiPMs) produced by Fondazione Bruno Kessler (FBK) \cite{FBK14}. The ongoing research and development on this detecting technology has led to the production of photosensors characterized by an array of Avalanche Photodiodes operating in Geiger mode (GM--APDs). They are based on a p$^+-~$n junction and are suitable to detect light in the blue and near-UV wavelength region \cite{Renker2006}. Moreover, if compared to Photomultiplier Tubes (PMTs), SiPMs present several advantages: compactness, ruggedness and insensitivity to magnetic fields. In addition, the breakdown voltage of these devices is low (order of 30~V) and shows a limited temperature dependence of about 24~mV/$^{\rm o}$C.
The Photo Detection Efficiency (PDE) values are higher than 25\%\ for 2~V of over--voltage (OV) and about 40\%\ at 6~V of OV in the range 390--410 nm \cite{Acerbi2015}. 

The tested  devices are NUV SiPMs of 1x1 mm$^2$ area with a cell of 40~$\mu$m and 50~$\mu$m and matrix of 16 SiPMs each of 3x3 mm$^2$ area and 40~$\mu$m cell. In addition, some preliminary results have been obtained testing a NUV SiPM characterized by a cell of 40~$\mu$m and an area of 6x6 mm$^2$.  

\subsection{Front--end Electronics}

\begin{figure}[ht!]
\centering
\includegraphics[height=0.35\textwidth,bb=0 0 323 165,clip]{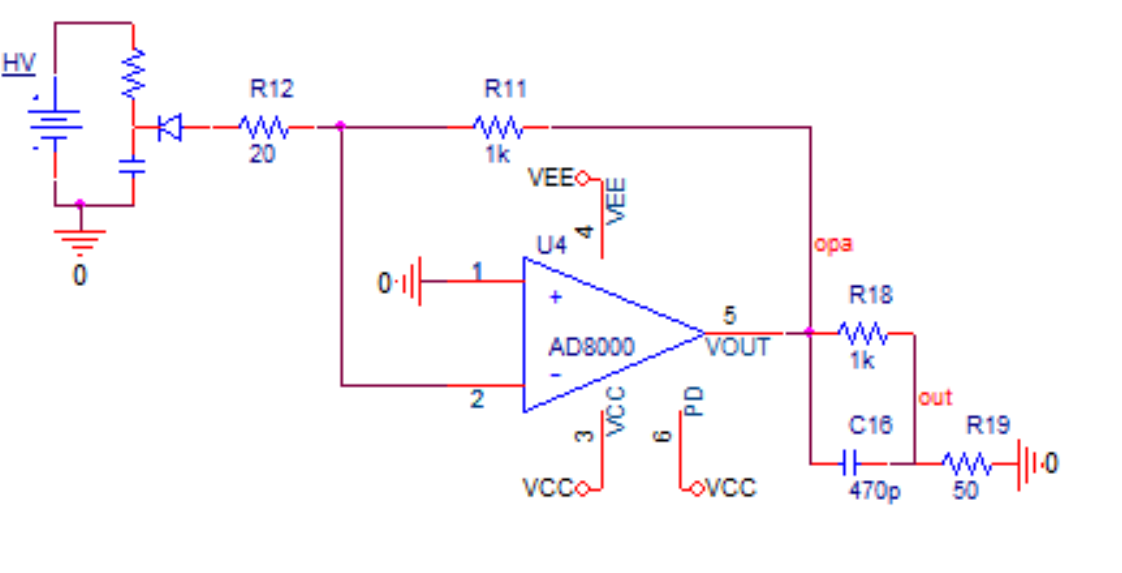} 
\caption{Scheme of the SiPM and the AD8000 OPA with a Pole-Zero cancellation circuit.}
\label{Fig_2}
\end{figure}

The front--end electronics plays a crucial role in the acquisition of signals produced by SiPMs. In order to optimize the current signal processing, a preamplifier has been designed based on a AD8000 operational amplifier (OPA) in a transimpedance configuration, as shown in Figure \ref{Fig_2}. A feedback resistance of 1~k$\Omega$ is used together with a 20~$\Omega$ decoupling resistance and the signal is sent to the preamplifier in DC mode. 

\begin{figure}[hb!]
\centering
\begin{tabular}{cc}
\includegraphics[height=0.24\textheight,bb=0 0 919 715,clip]{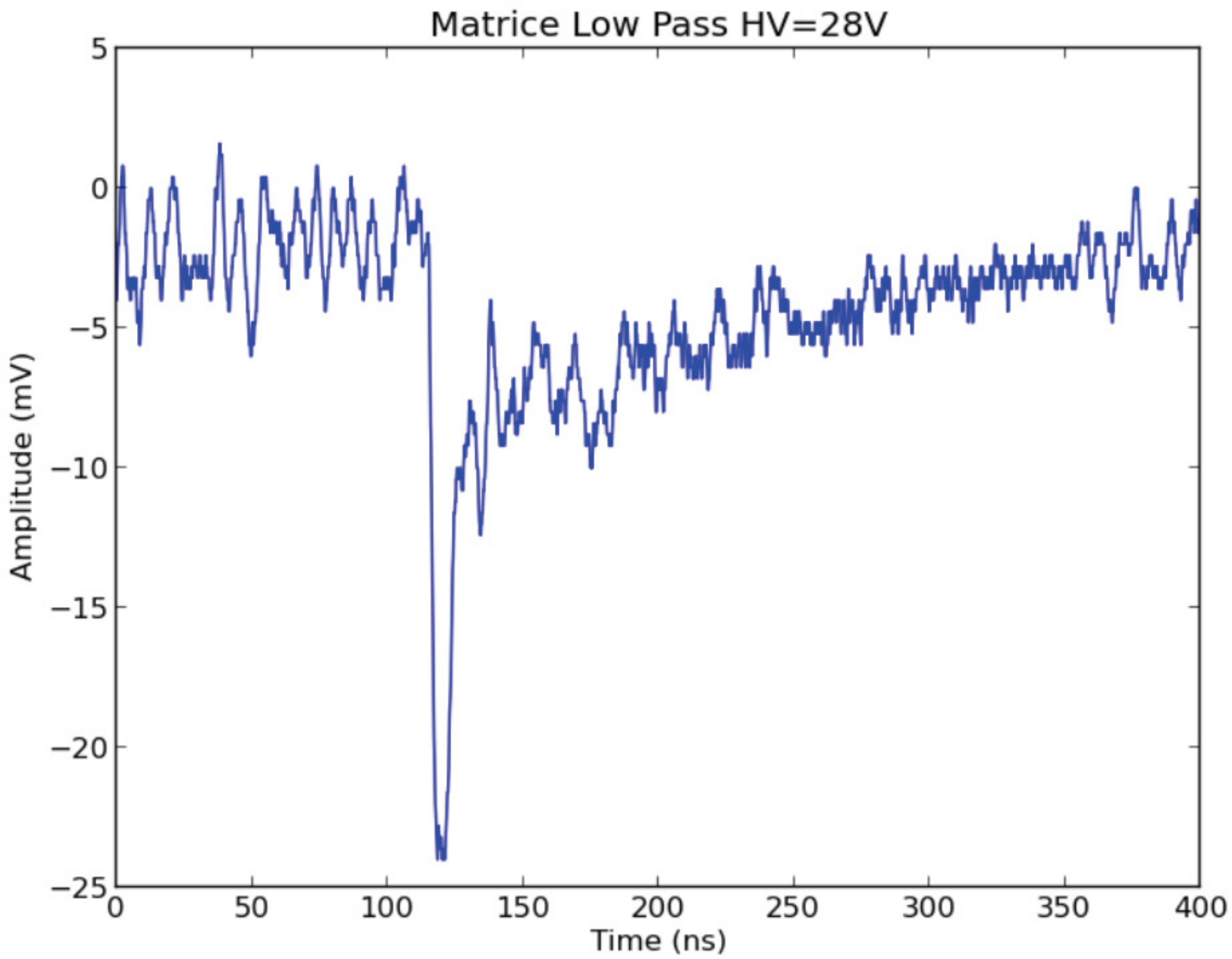} &
\includegraphics[height=0.24\textheight,bb=0 0 708 550,clip]{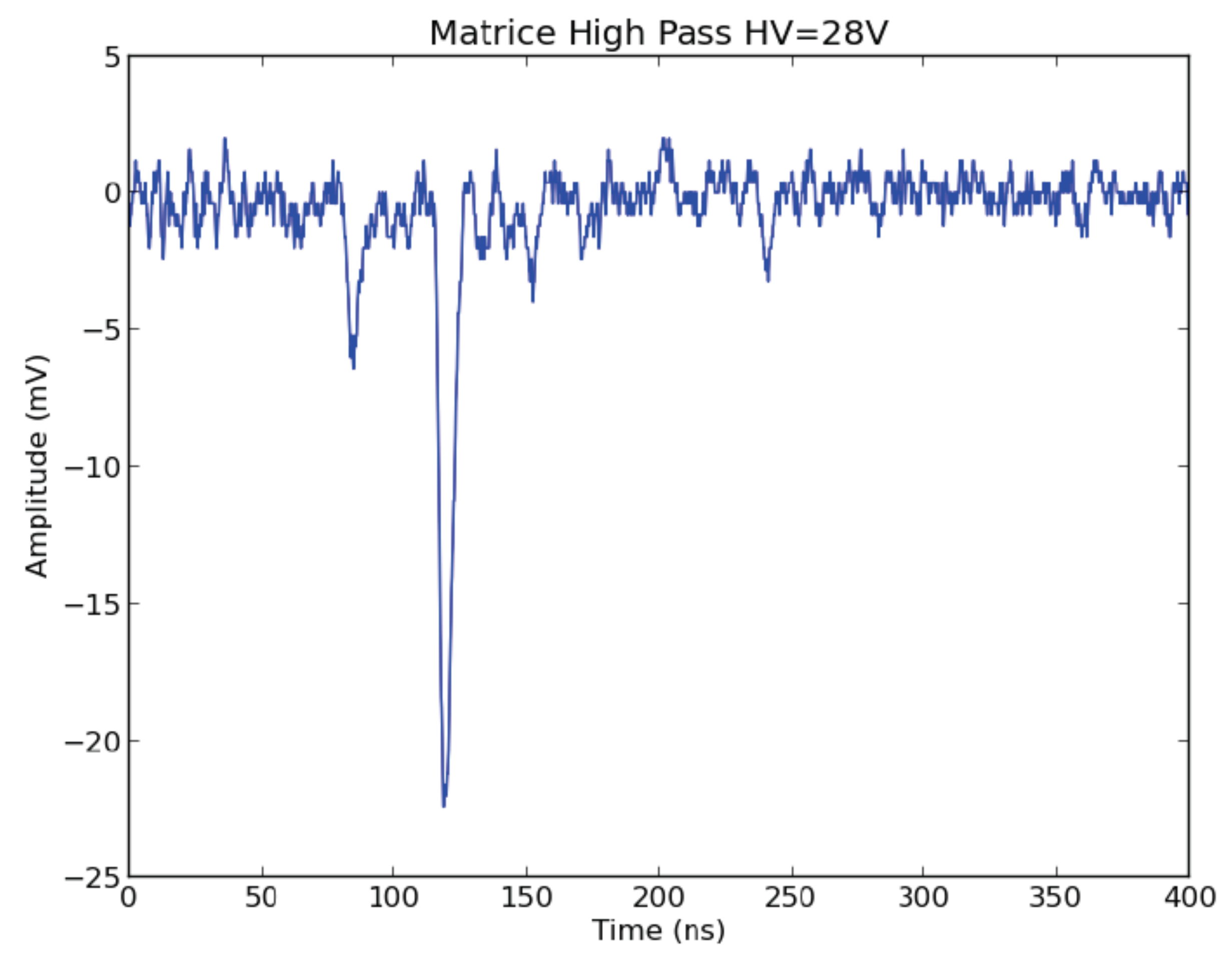}
\end{tabular}
\caption{Current signals from the preamplifier collected
without ({\it left panel}) and with the Pole-Zero
cancellation circuit ({\it right panel}).}
\label{Fig_3}
\end{figure}

The original design has been modified by adding a Pole--Zero cancellation circuit  \cite{GOL13} in order to get rid of the long tail which characterizes the signal as shown in the left panel of Figure \ref{Fig_3} . The origin of this effect is the signal recovery time due to the single cell capacitance and the quenching resistor.
The correction of the recovery time avoids the pile--up and the consequent loss of information in the signal. The filter has been optimized with a capacitance of 470~pF in parallel to a 1~k$\Omega$ resistance with a load of 50~$\Omega$ (Figure \ref{Fig_2}). As shown in the right panel of Figure \ref{Fig_3}, the net effect of the Pole-Zero cancellation circuit is to attenuate the long tail with a minimal effect on the signal amplitude. Further information about the development of the readout system can be found in \cite{Ambrosi}.

\section{Silicon Photomultipliers Performances}
\begin{figure}[t!]
\centering
\begin{tabular}{cc}
\includegraphics[height=0.2\textheight,bb=0 0 351 235,clip]{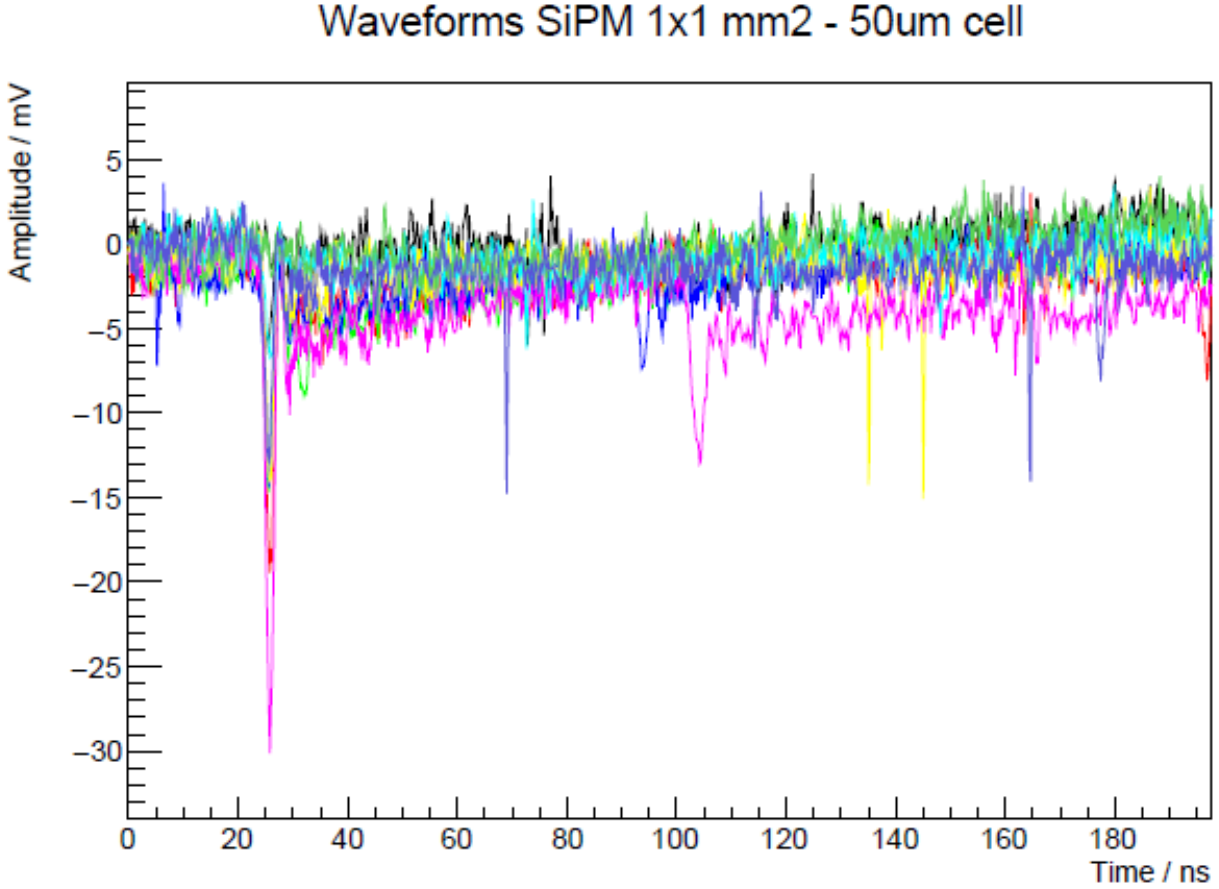} &
\includegraphics[height=0.2\textheight,bb=9 9 409 270,clip]{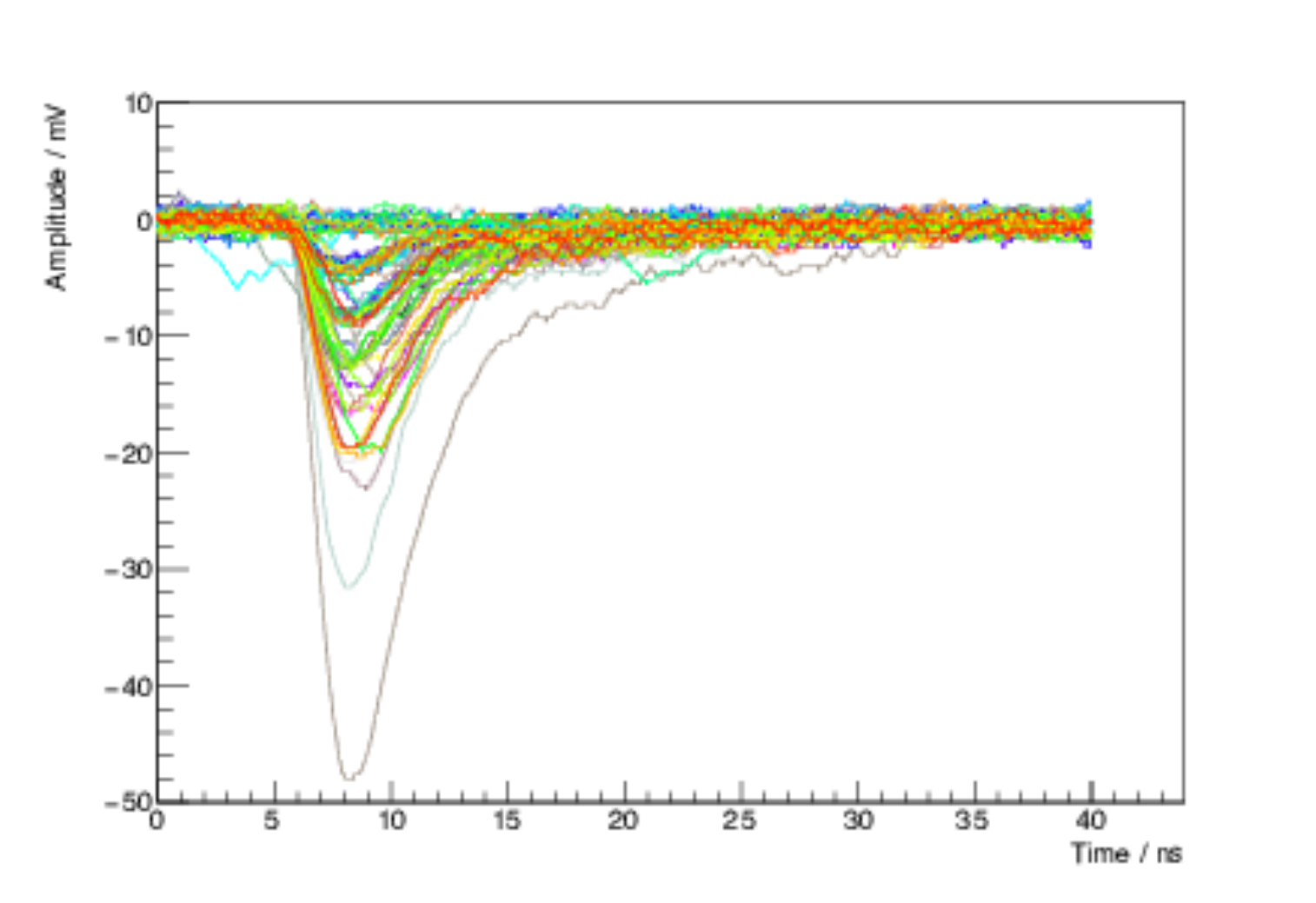} \\
\multicolumn{2}{c}{
\includegraphics[height=0.2\textheight,bb=5 9 550 350,clip]{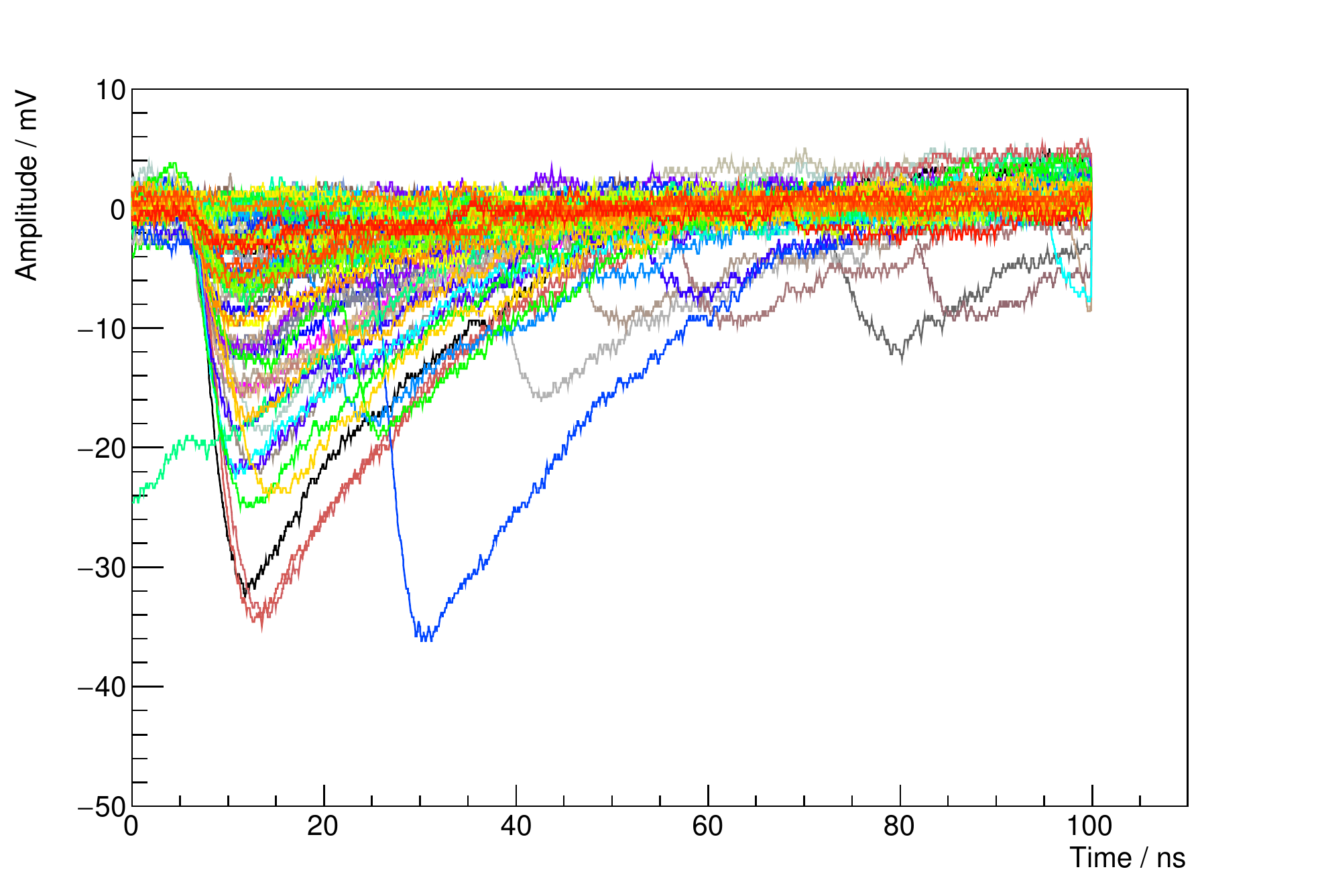}}
\end{tabular}
\caption{Amplitude spectra:
{\it Top left panel}: 1x1 mm$^2$ SiPM;
{\it Top right panel}: 3x3 mm$^2$ SiPM;
{\it Bottom panel}: 6x6 mm$^2$ SiPM.}
\label{waveforms}
\end{figure}

The laboratory tests have been performed in a dark box where a laser (emission at 380 nm), the photodetector, the electronics and a system of thermocouples have been placed. The original design of the amplifier circuit was coupled to the SiPMs of 1$\times$1~mm$^2$ area (both 40 $\mu$m and 50 $\mu$m cell), while the one modified with the filter was used for waveform acquisition of the 3$\times$3 mm$^2$ and 6$\times$6 mm$^2$ SiPMs. In the latter configuration, a printed circuit board (PCB) equipped with 16 AD8000 OPAs collected the current signals from the SiPMs and it was, in turn, connected to another PCB endowed with 16 LEMO cables for the signal readout. These were sent to a TD5S5104B oscilloscope in the case of the 3$\times$3 mm$^2$ and 6$\times$6 mm$^2$ SiPMs or to a DT5742 Switched Capacitor Digitizer (12 bit, 5 GS/s - 1024 storage cells/CH) for the 1$\times$1 mm$^2$ SiPM. 
As can been observed in Figure \ref{waveforms}, SiPMs of different  area showed different signal widths (from 10 ns to 40 ns). The effect of the Pole--Zero cancellation network is evident for the signals collected from the 3$\times$3 mm$^2$ detectors for which the filter has been optimized. 

\renewcommand{\tabcolsep}{2pt}

\subsection{1x1 mm$^2$ Area SiPMs}

\begin{figure}[p!]
\centering
\begin{tabular}{cc}
\includegraphics[height=0.18\textheight,bb=20 9 500 340,clip]{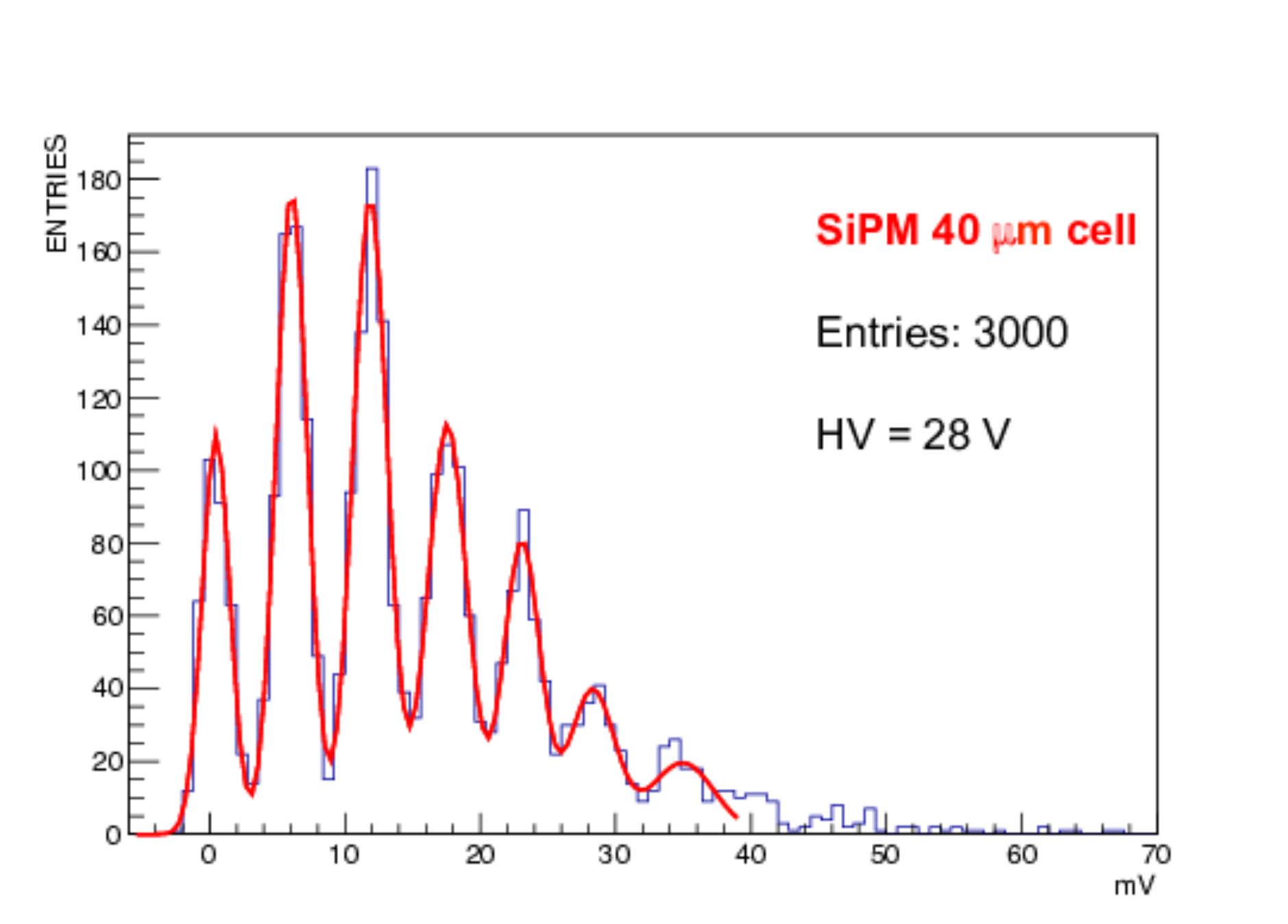} &
\includegraphics[height=0.18\textheight,bb=15 9 520 355,clip]{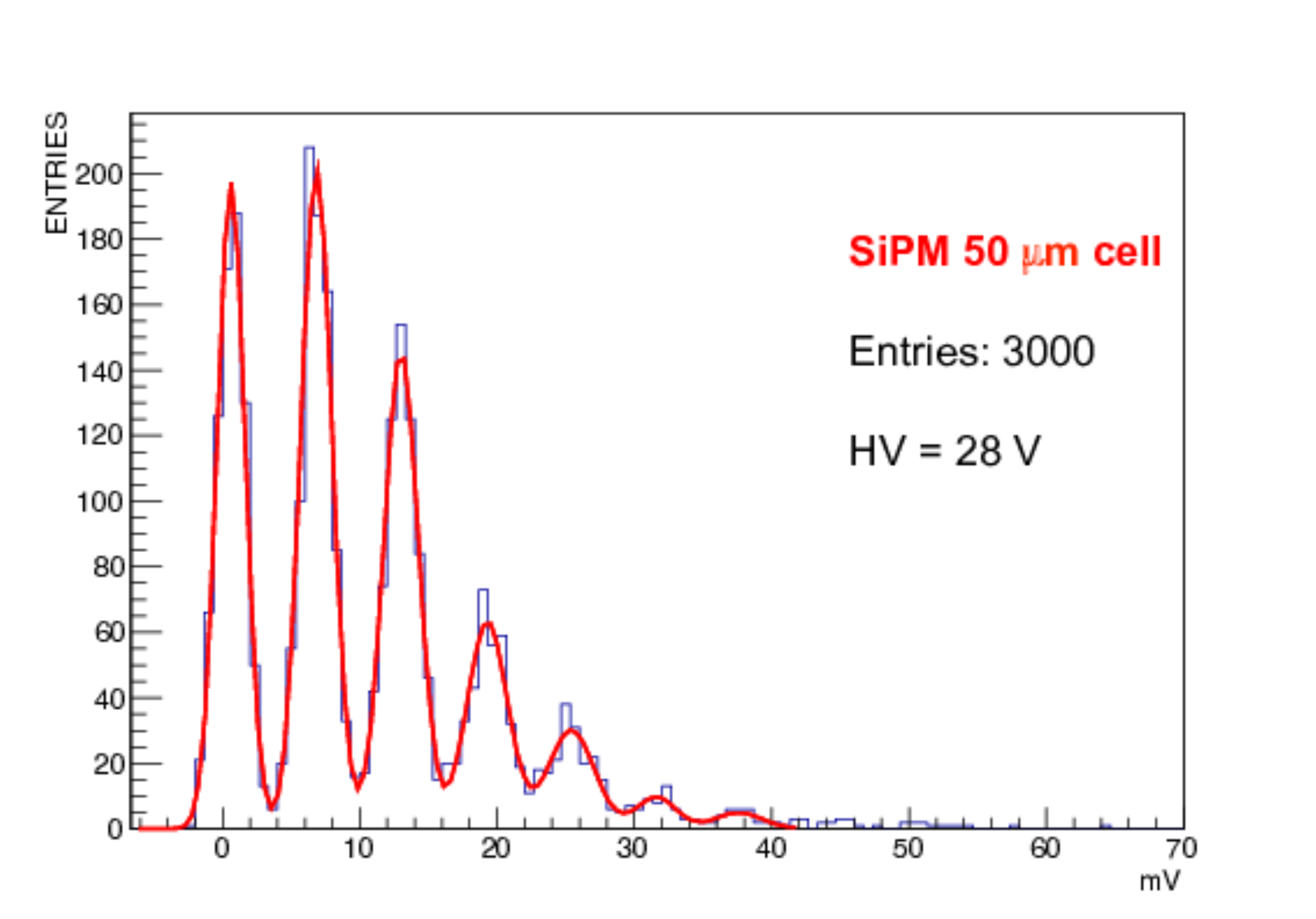}
\end{tabular}
\caption{Amplitude spectrum and multi--Gaussian fit.
{\it Left panel}: 40 $\mu$m cell SiPM ;
{\it Right panel}: 50 $\mu$m cell SiPM.}
\label{manine}
\vspace{12pt}
\begin{tabular}{cc}
\includegraphics[height=0.18\textheight,bb=10 10 502 332,clip]{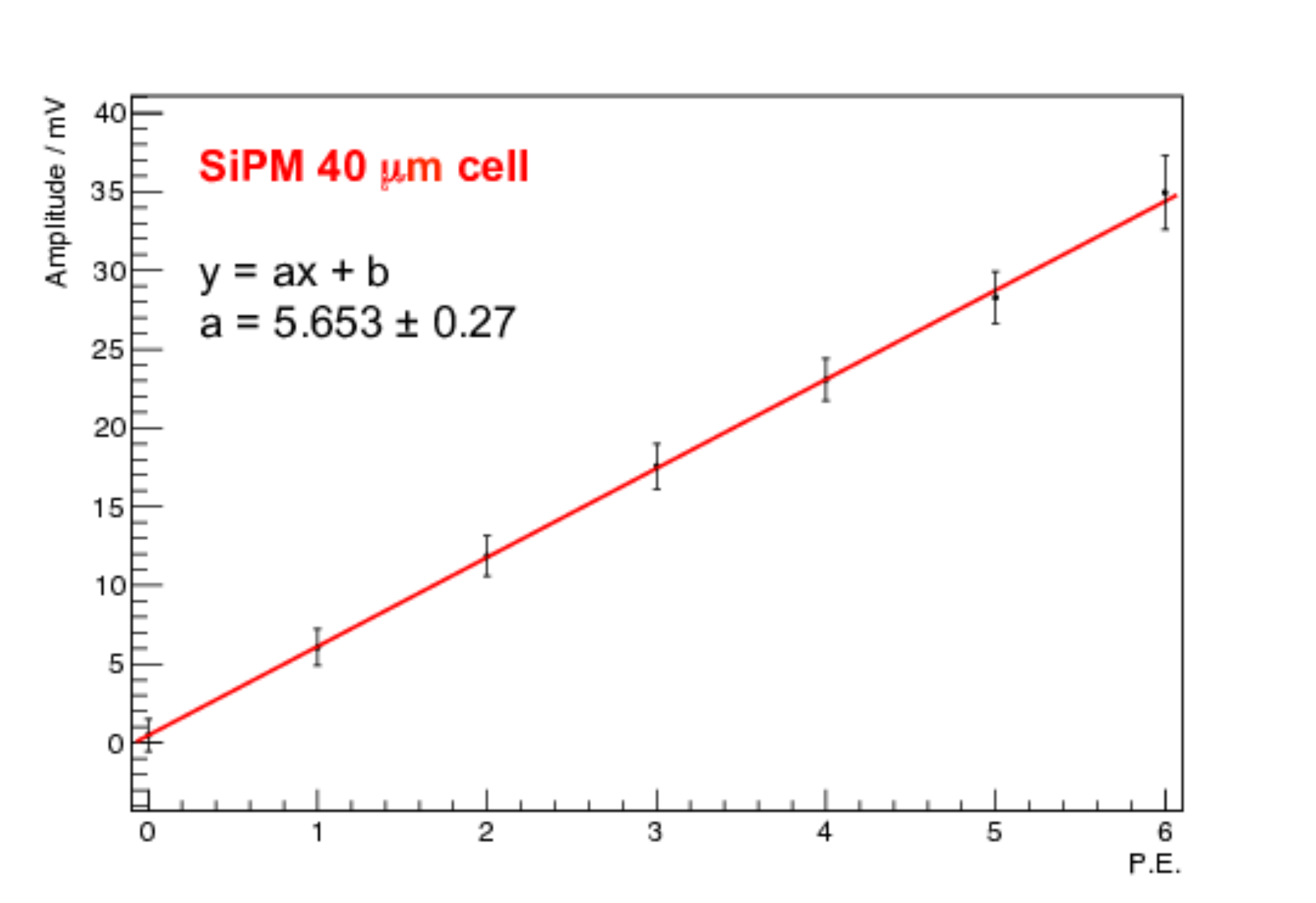} &
\includegraphics[height=0.18\textheight,bb=20 10 532 348,clip]{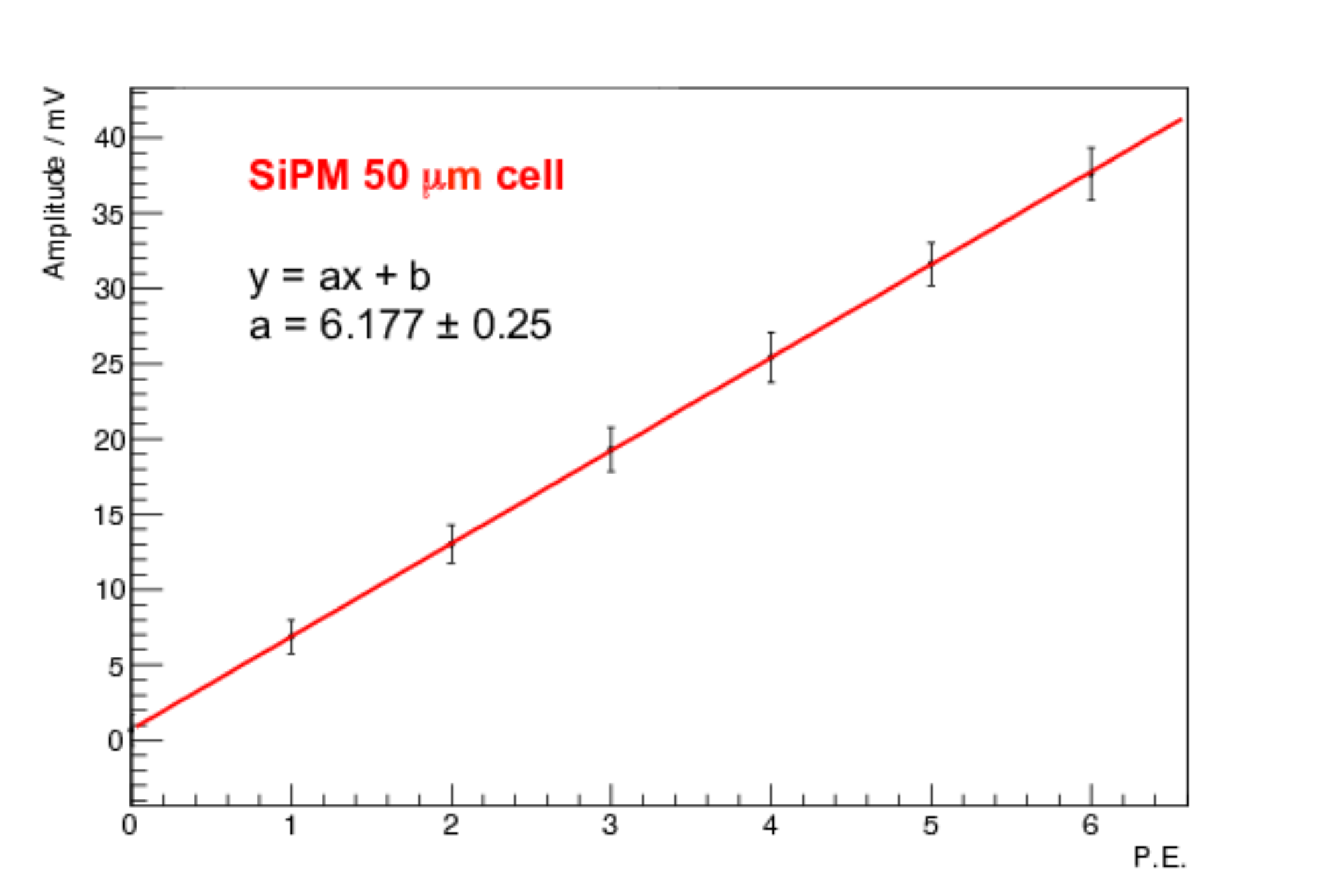}
\end{tabular}
\caption{Amplitude as a function of the number of photoelectrons (PE).
{\it Left panel}: 40 $\mu$m cell SiPM ;
{\it Right panel}: 50 $\mu$m cell SiPM.}
\label{gain_ampl}
\vspace{12pt}
\begin{tabular}{cc}
\includegraphics[height=0.18\textheight,bb=10 10 520 360,clip]{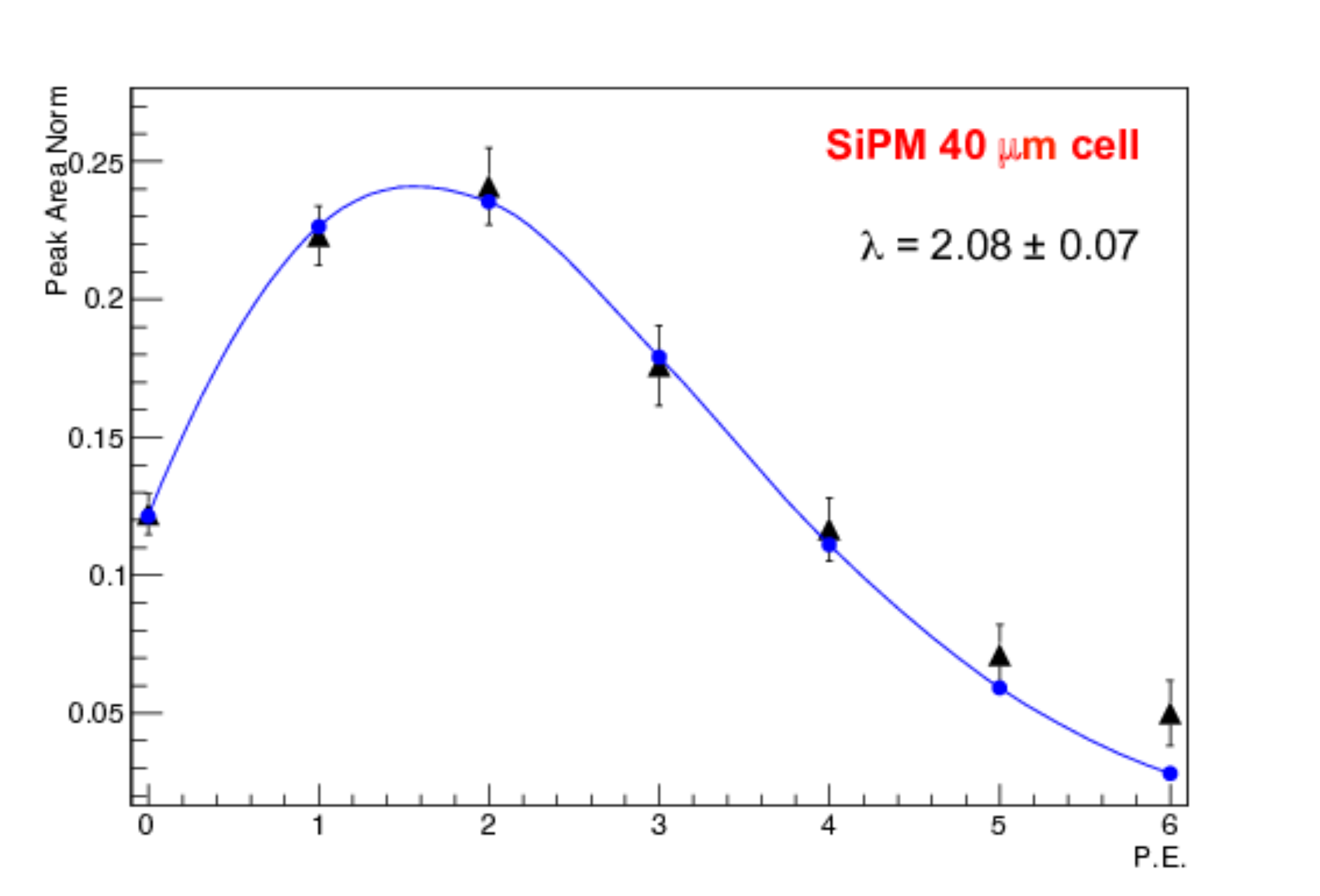} &
\includegraphics[height=0.18\textheight,bb=10 10 520 360,clip]{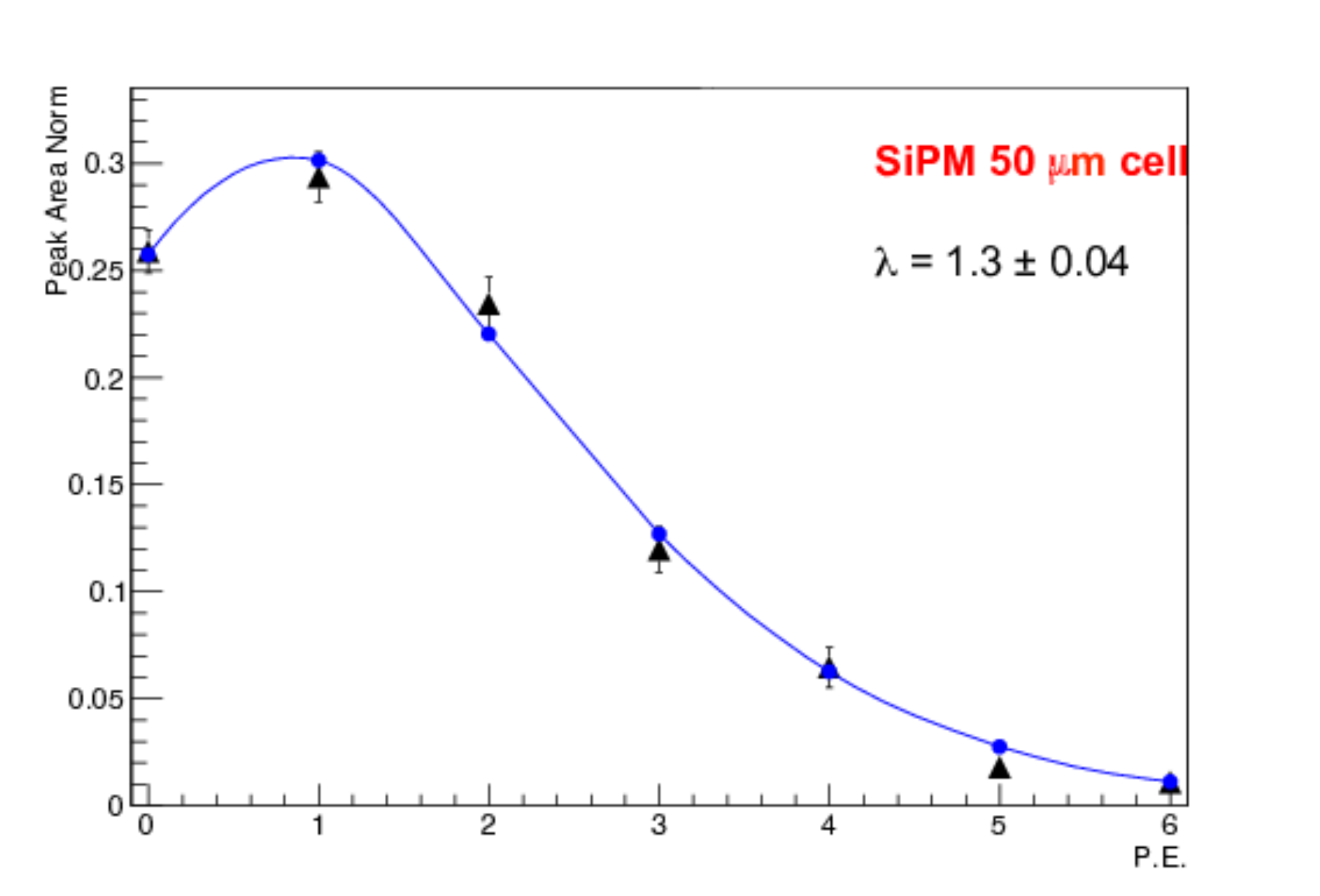}
\end{tabular}
\caption{Amplitude distribution with Poisson fit superimposed. Number of photons ($\lambda$) is also shown.
{\it Left panel}: 40 $\mu$m cell SiPM ;
{\it Right panel}: 50 $\mu$m cell SiPM.}
\label{poisson}
\vspace{12pt}
\begin{tabular}{cc}
\includegraphics[height=0.18\textheight,bb=20 5 515 350,clip]{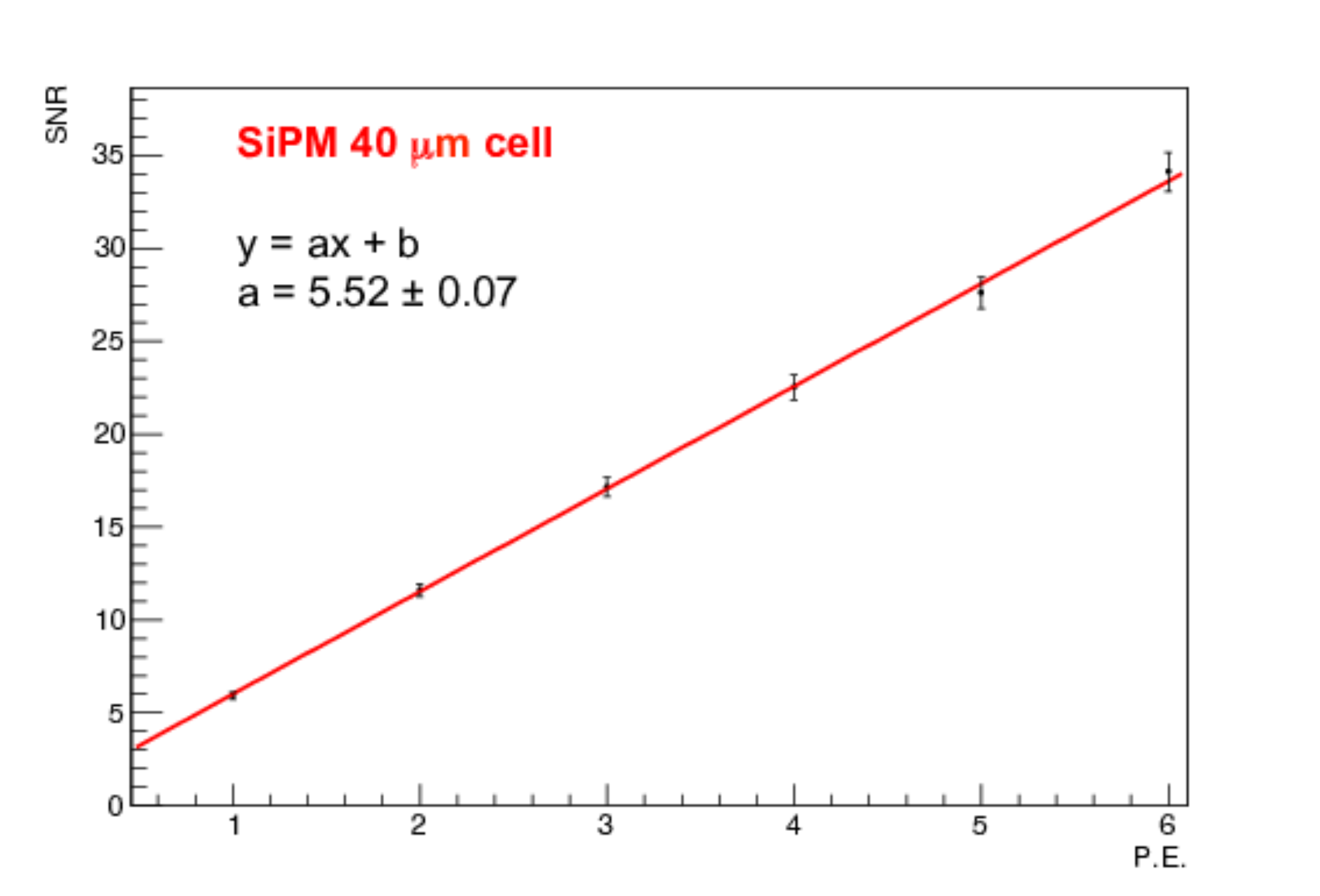} &
\includegraphics[height=0.18\textheight,bb=20 5 573 350,clip]{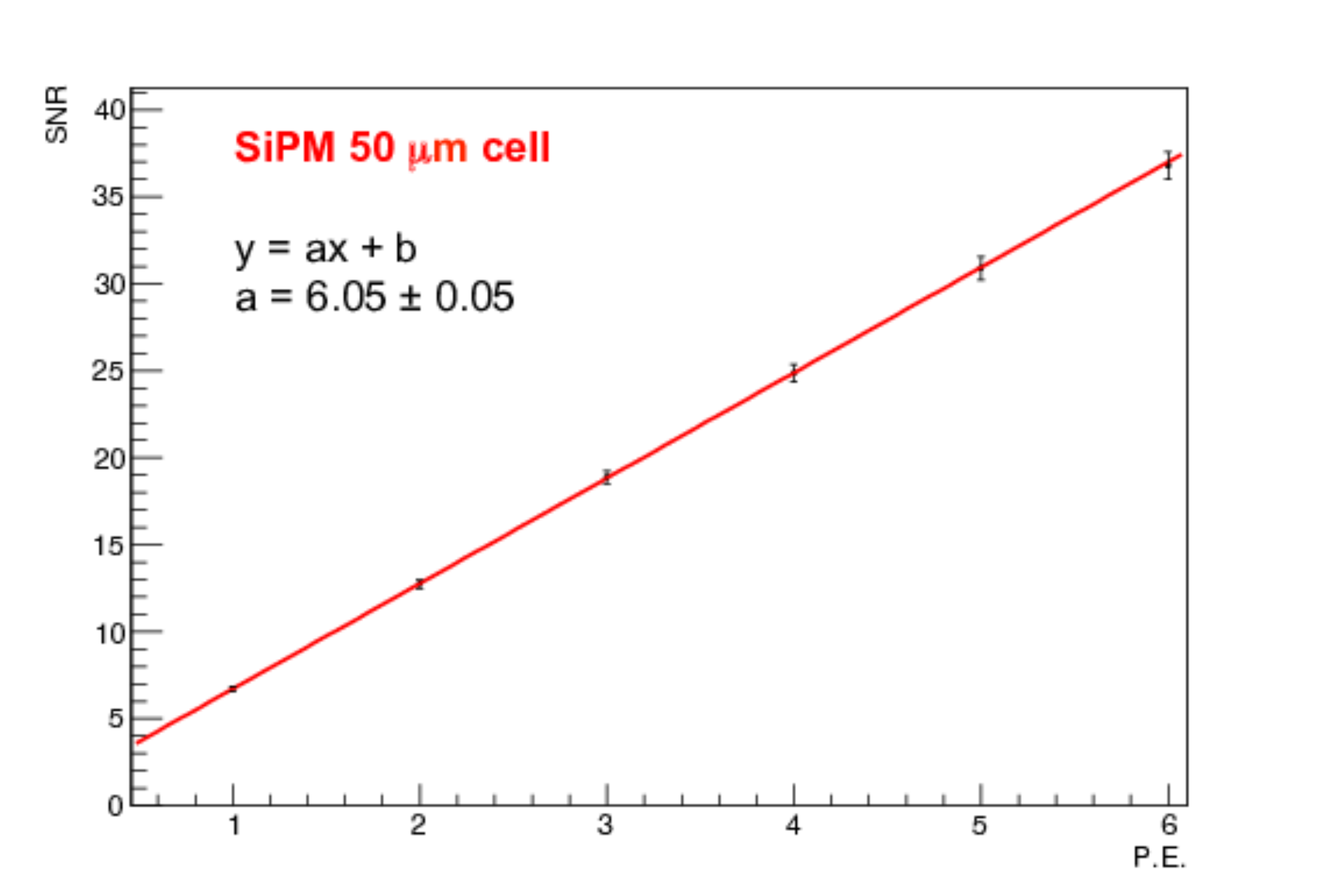}
\end{tabular}
\caption{Signal to noise ratio (SNR).
{\it Left panel}: 40 $\mu$m cell SiPM ;
{\it Right panel}: 50 $\mu$m cell SiPM.}
\label{SNR_1x1}
\end{figure}

\begin{figure}[ht!]
\centering
\includegraphics[height=0.23\textheight,bb=10 10 420 300,clip]{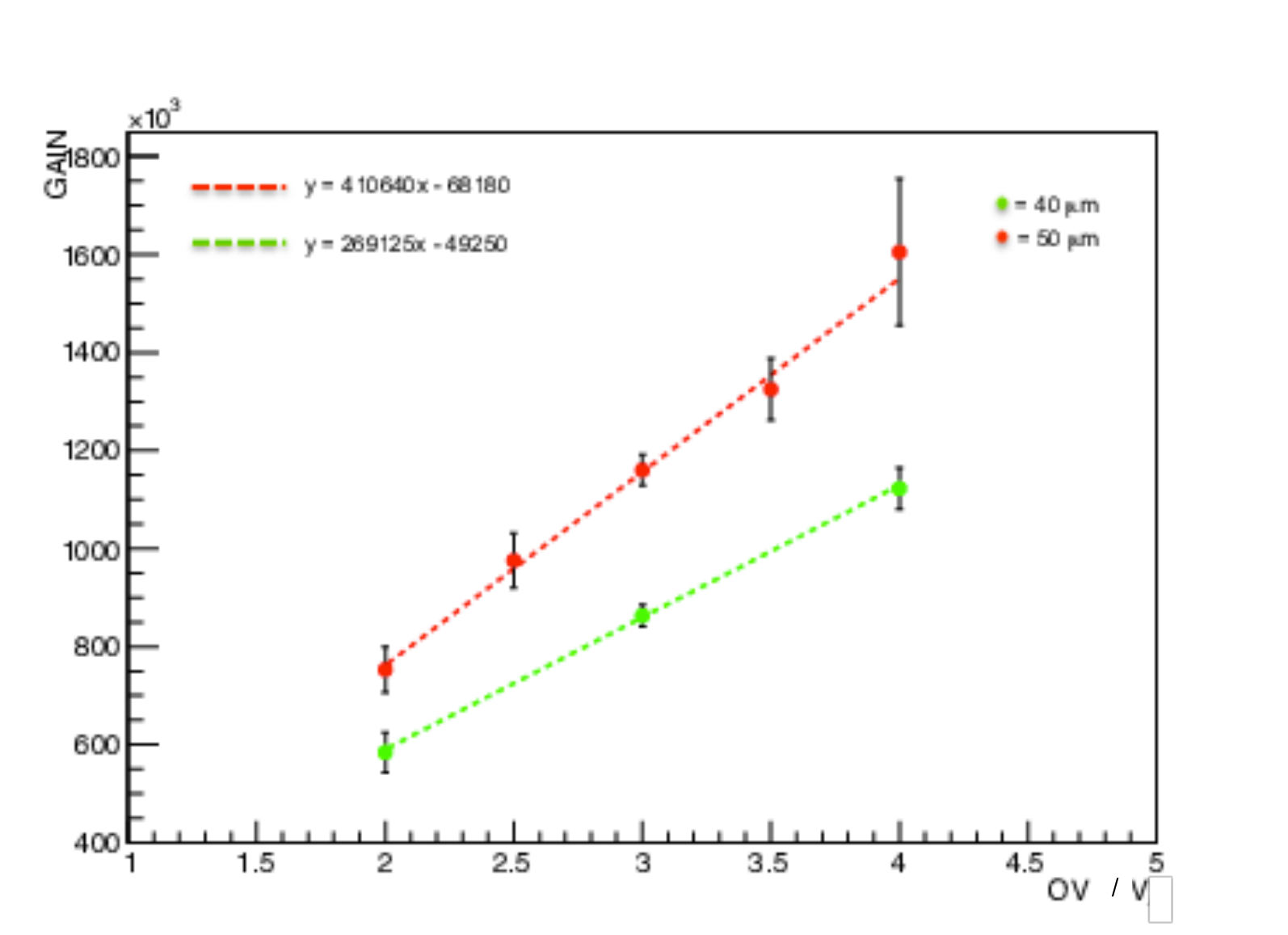} 
\caption{Comparison between the gain of 40 $\mu$m cell SiPM and 50 $\mu$m cell SiPMs.}
\label{gain_carriers}
\end{figure}

Measurements have been performed on 1$\times$1 mm$^2$ area SiPMs with cell size of 40~$\mu$m and 50~$\mu$m from a previous production run.
The amplitude spectrum relative to the devices is represented in Figure \ref{manine} and the amplitude vs.~the converted photoelectron number is illustrated in Figure \ref{gain_ampl} showing the expected linear correlation.
Figure \ref{poisson}  represents the Poissonian distribution exhibited by the amplitude distribution, which allows to estimate the number of photons detected by the SiPM. The signal-to-noise ratio (SNR) has been evaluated as the ratio between the amplitude of the single photoelectron peaks to the width of the electronic noise, as illustrated in Figure \ref{SNR_1x1}.
Moreover, the gain for different overvoltages has been  evaluated for both the devices. This has been calculated by integrating the charge produced by the SiPMs and performing a multi--Gaussian fit to the charge distribution.
It is interesting to observe that the 50 $\mu$m cell detector is characterized by a higher gain with respect to the 40 $\mu$m cell detector, for the same overvoltage value as shown in Figure \ref{gain_carriers}. Nevertheless at the same gain value, the SNR is higher for the 40 $\mu$m cell SiPM compared to the 50 $\mu$m cell detector. 

\FloatBarrier

\begin{figure}[ht!]
\centering
\includegraphics[height=0.4\textwidth,bb=10 0 380 230,clip]{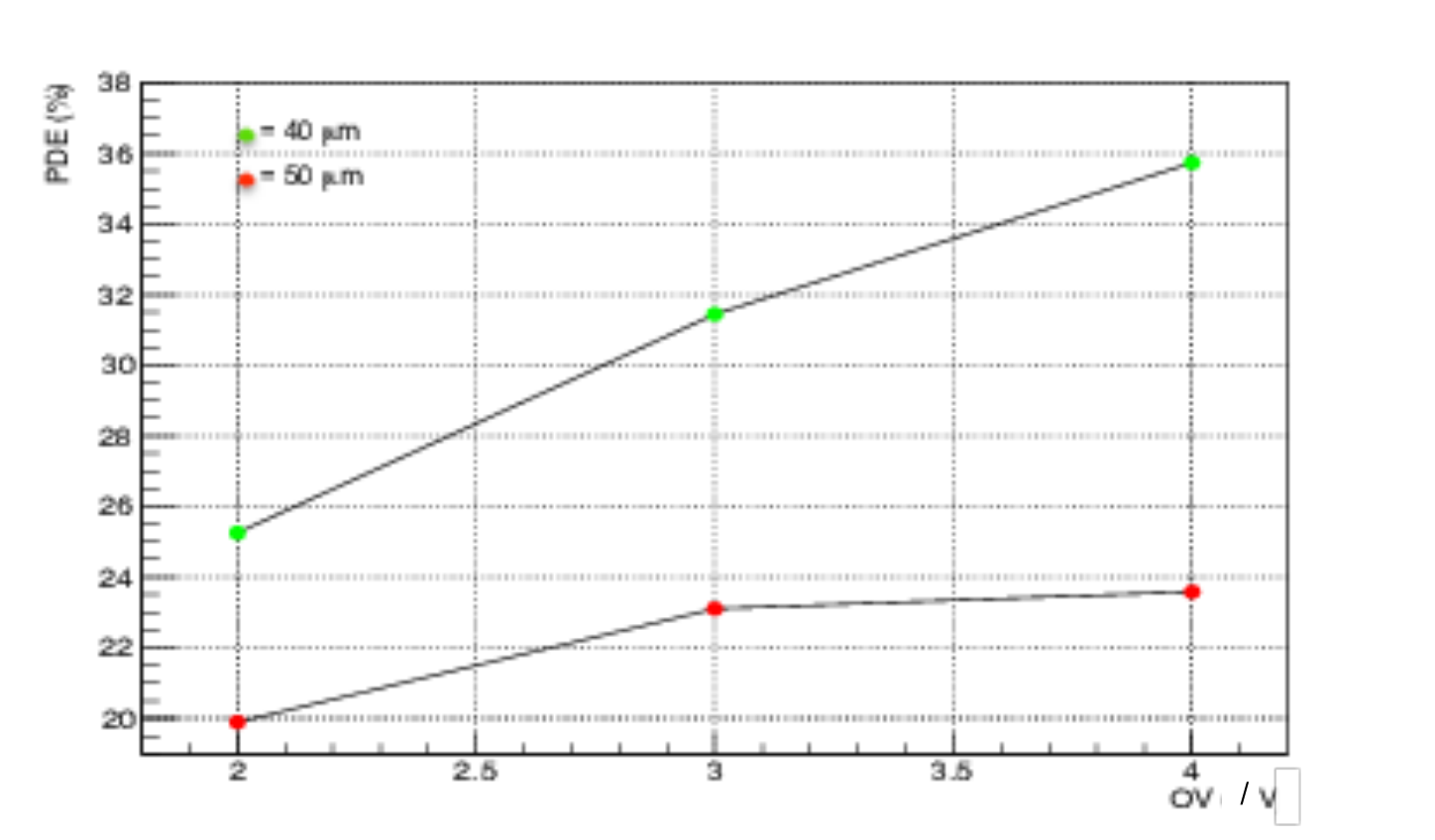} 
\caption{Photo Detection Efficiency as a function of the excess bias (for a measurement at T = 21$^{\rm o}$ C).}
\label{PDE_vs_OV}
\end{figure}

\begin{figure}[ht!]
\centering
\begin{tabular}{cc}
\includegraphics[height=0.21\textheight,bb=0 0 376 258,clip]{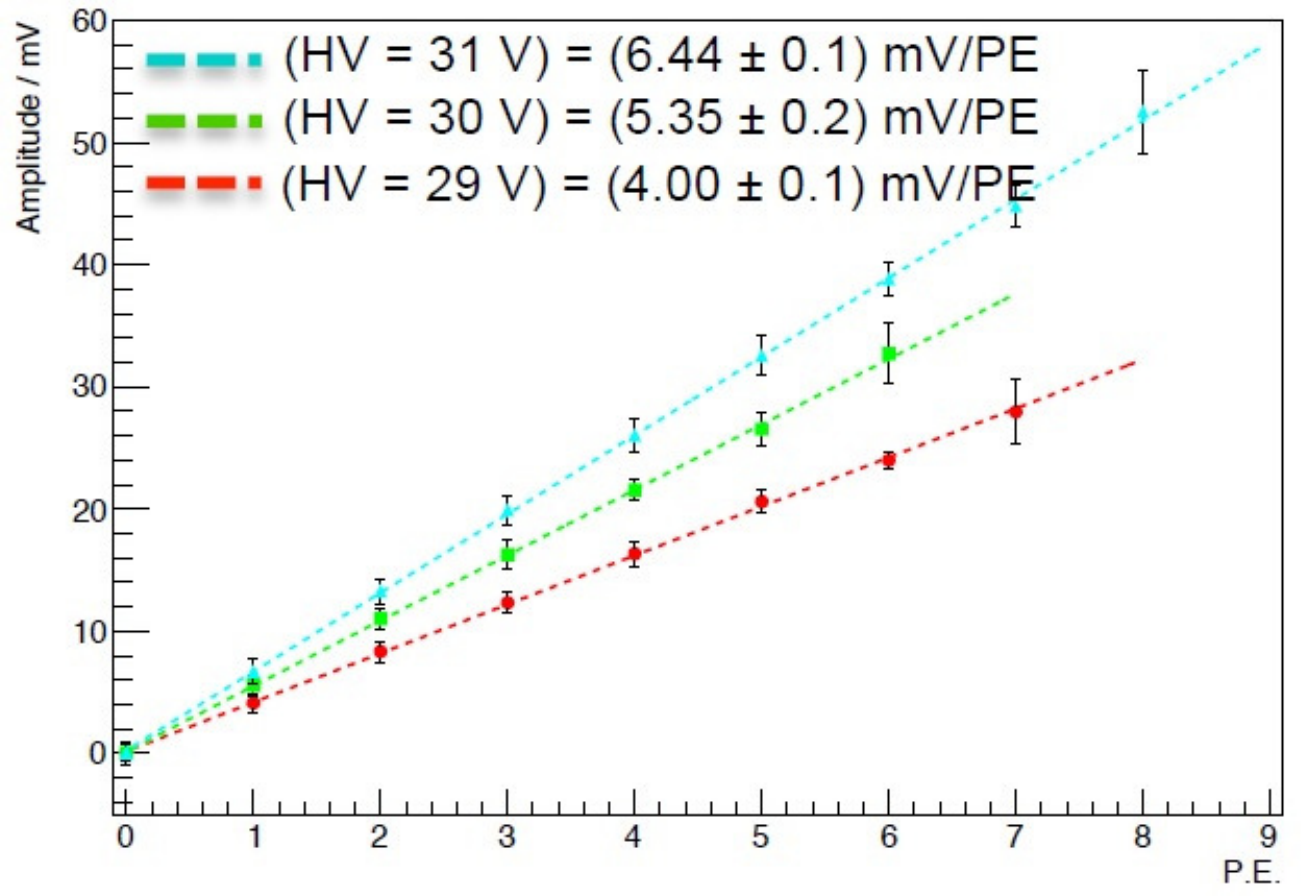} &
\includegraphics[height=0.23\textheight,bb=0 0 383 266,clip]{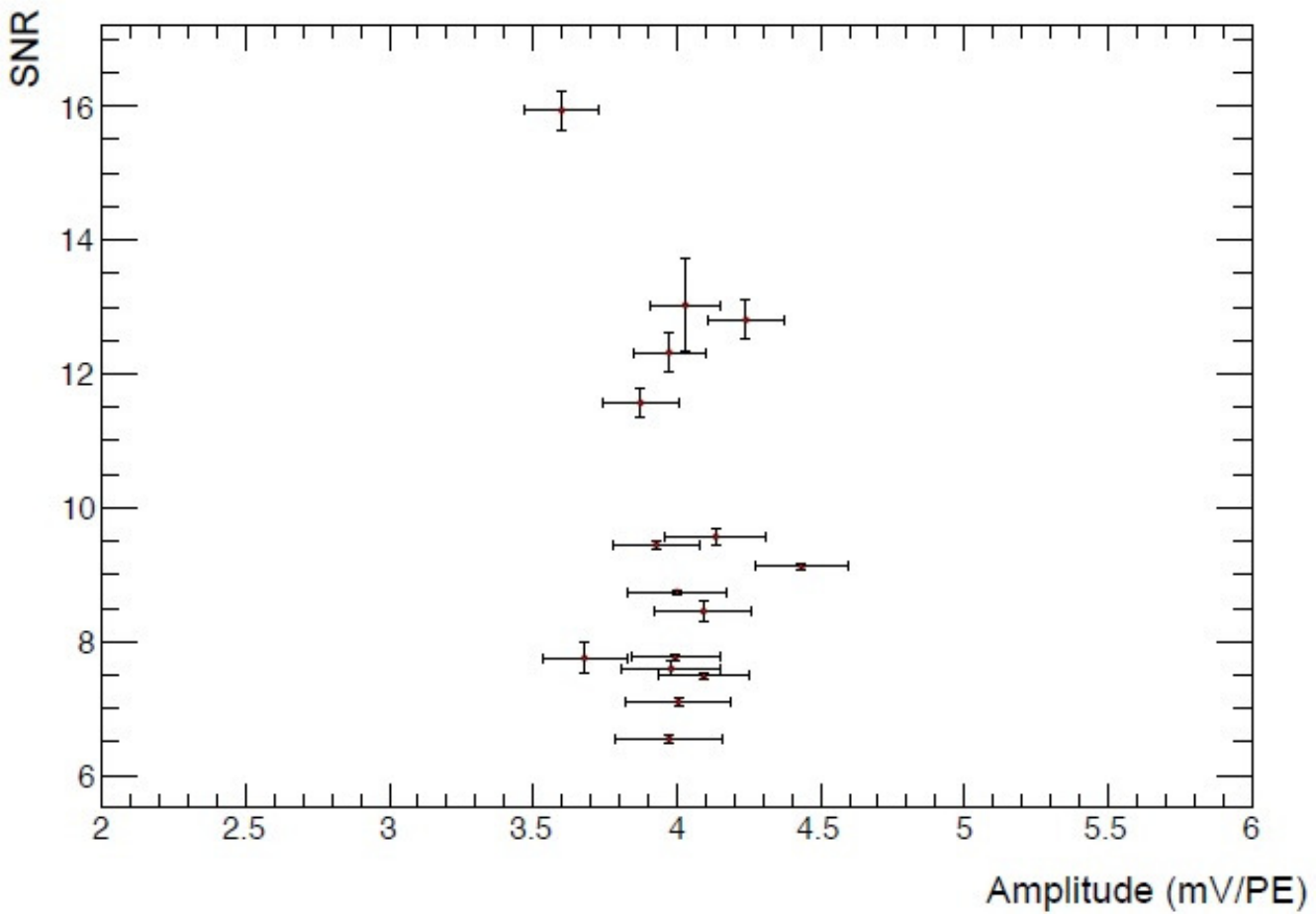} \\
\includegraphics[height=0.20\textheight,bb=0 10 450 270,clip]{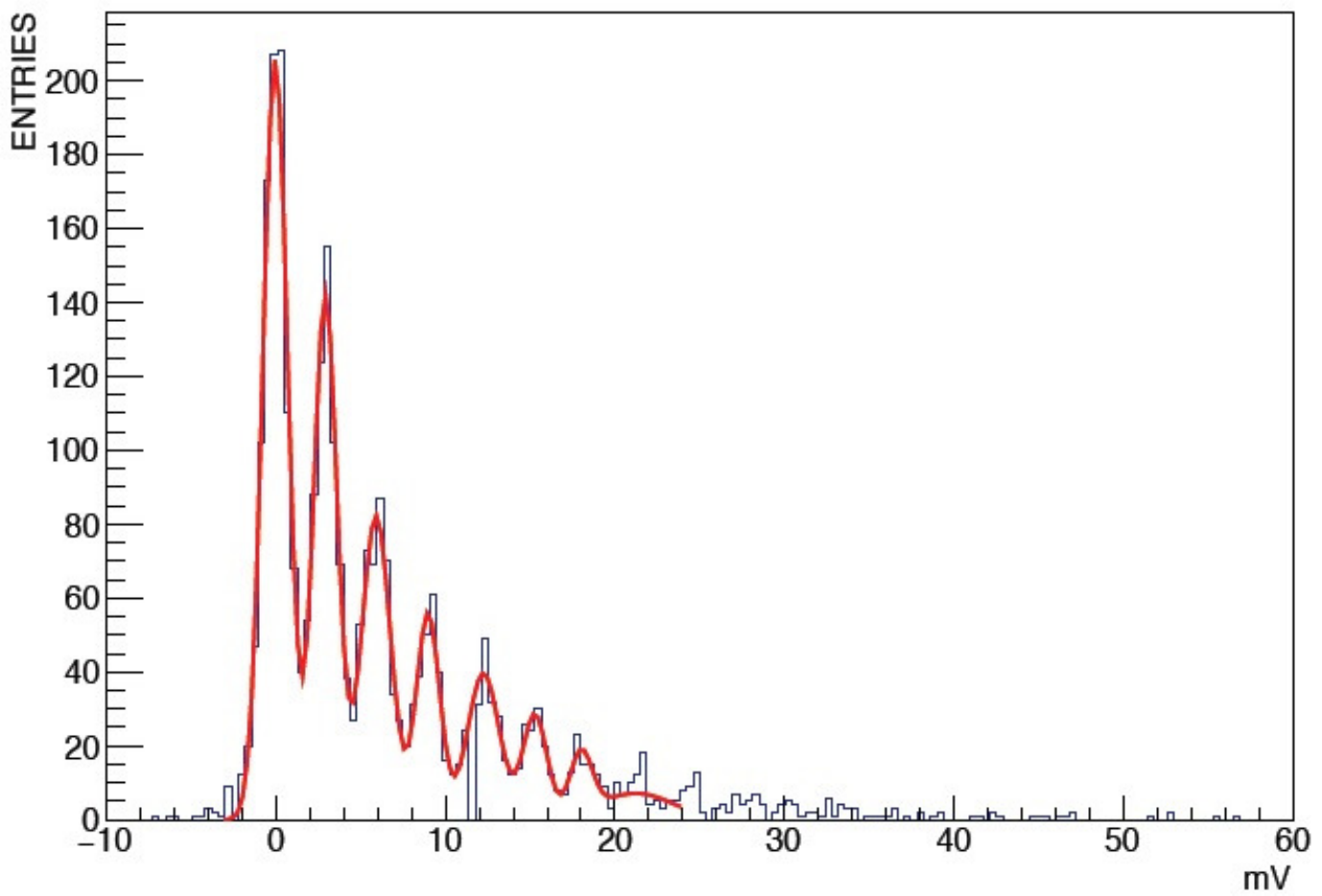} &
\includegraphics[height=0.23\textheight,bb=0 0 318 237,clip]{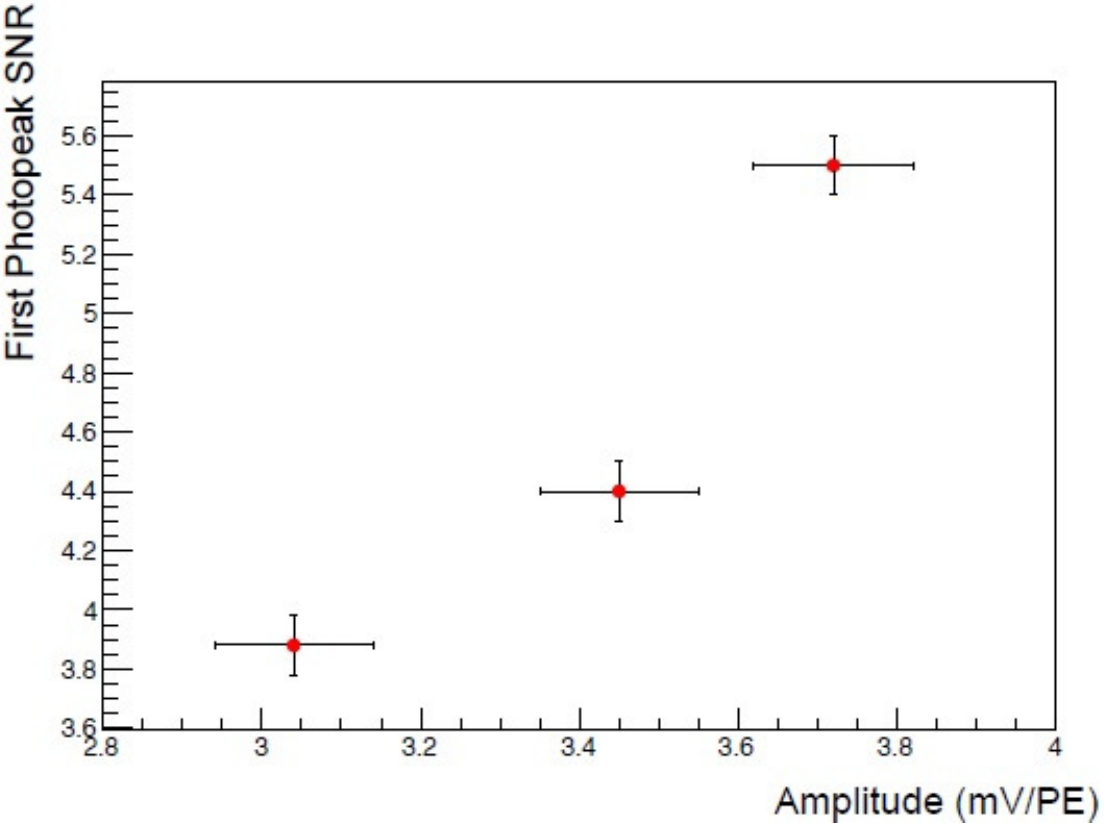}  
\end{tabular}
\caption{3$\times$3 mm$^2$ SiPM (top panels) and 6$\times$6 mm$^2$ SiPM (bottom panels).
{\it Top left panel}: Amplitude distribution as function of photoelectron number (3$\times$3 mm$^2$ SiPM) ;
{\it Top right panel}: Signal to noise ratio for the 16 SiPMs of the matrix ;
{\it Bottom left panel}: Amplitude spectrum (6$\times$6 mm$^2$ SiPM) ;
{\it Bottom right panel}: Signal to noise ratio (6$\times$6 mm$^2$ SiPM).}
\label{fig_3x3_6x6}
\end{figure}

The PDE has been estimated,  as the ratio between the number of photons detected by the SiPM to the number of photons detected by a calibrated photodiode (Hamamatsu s3759), using as light source the laser in pulsed-mode. The measurements have been performed with an integrating sphere equipped with three ports on which the laser head and the detectors have been placed.
The PDE as a function of the overvoltage is shown in Figure \ref{PDE_vs_OV} and the measured values confirm the specification given by FBK.

\subsection{3x3 mm$^2$ Area SiPMs Matrix}

The measurements of current signals from the SiPM sixteen--element matrix, where each detector has an area of 3$\times$3 mm$^2$, have been performed by covering all but one channel with a black mask.
Three operating voltages corresponding to 3, 4 and 5 V of overvoltage  have been used, verifying the dependence of the signal amplitude with the bias. As shown in Figure \ref{fig_3x3_6x6} (top left panel) the values fall between 4.00 mV/PE and 6.44 mV/PE.
It is interesting to note that the amplitude shows a uniformity better than 10\% and the variation in SNR is limited to few \%.
This performance makes the matrix of 16 SiPMs coupled to the above--described electronics a promising detecting system for applications requiring a complex structure, such as the SiPM based camera of CTA. 

\subsection{6x6 mm$^2$ Area SiPMs}

Preliminary measurements have been performed on SiPMs with 6$\times$6 mm$^2$ area and a cell size of 40 $\mu$m.
These SiPMs belong to the last production of FBK, have a breakdown voltage of about 32 V and a geometry optimized to reduce the dead zone area between the SiPMs when mounted in matrices.

An array of 4 SiPMs has been considered and, as previously described, for each run only one detector has been exposed to the light by covering the other three. Three biases have been selected (overvoltage of 6, 7 and 8 V) and an example of amplitude spectrum taken at 6 OV is shown in the bottom left panel of Figure \ref{fig_3x3_6x6}.
The Poissonian fit of the amplitude provides a value for the cross talk of 38\%. Finally, we evaluated the first photopeak signal--to--noise ratio and, as illustrated in the bottom right panel of Figure \ref{fig_3x3_6x6}, we observed that saturation is not yet reached.
The SNR could be improved by increasing the amplitude (and thus the operating voltage) but a compromise must be found in order to avoid fluctuations in the signal due to electronic noise.

\section{Future Work}

\begin{figure}[ht!]
\centering
\includegraphics[height=0.35\textwidth,bb=0 0 150 165,clip]{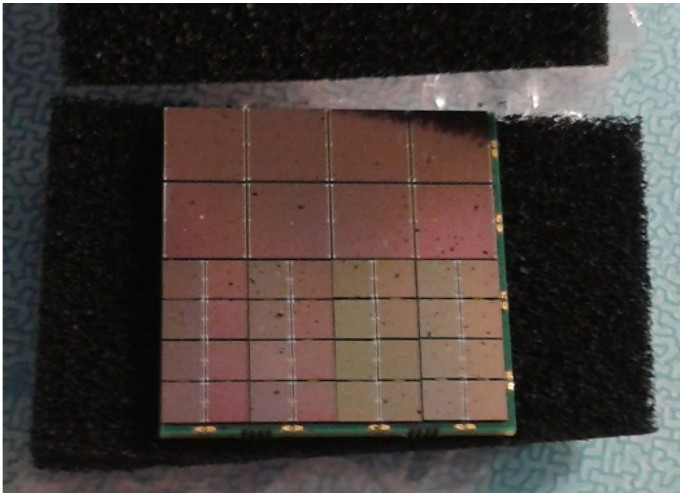} 
\caption{Array of SiPMs containing detectors of different areas.}
\label{new_matrix}
\end{figure}

A new generation of NUV SiPM with small cell size is under production at FBK. Based on measurement on prototypes we do expect a geometrical fill factor in excess of 75\% on the single 6$\times$6 mm$^2$ devices, a PDE of about 40\% and low correlated noise.
In parallel,  the configuration shown in Figure  \ref{new_matrix} assembled with 3$\times$3 mm$^2$ SiPMs from two different productions and the new 6$\times$6 mm$^2$ is already under investigation and it will be useful to better compare the performances between devices of different areas.

At the same time, further studies are in progress to improve the front--end electronics in terms of signal sampling, optimizing the noise from diffuse light, and reducing the current signal tail.  

\section{Acknowledgments}
We gratefully acknowledge support from the agencies 
and organizations 
listed under Funding Agencies at this website: 
{\tt http://www.cta-observatory.org/}.

\begin{thebibliography}{99}
\bibitem{FBK14} {\tt{http://www.fbk.eu/}}
\bibitem{Renker2006} D.~Renker,  {\it ``Geiger-mode avalanche photodiodes, history, properties and problems''}, {\bf Nucl.Instrum.Methods A},  Vol. 567, pp. 48-56, 2006
\bibitem{Acerbi2015} F.~Acerbi et al., {\it ``NUV Silicon Photomultipliers With High Detection Efficiency and Reduced Delayed Correlated-Noise''}, {\bf IEEE TRANSACTIONS ON NUCLEAR SCIENCE},  Volume: 62, Issue:3, pp.1318,1325, June 2015
\bibitem{PRO13} T.~Pro, et~al., {\it ``New Developments of Near-UV SiPMs at FBK''}, {\bf IEEE TRANSACTIONS ON NUCLEAR SCIENCE},  Volume: 60, Issue:3, Part:3 (2013)
\bibitem{GOL13} A.~Gola, C.~Piemonte, A.~Tarolli, 
{\it ``Analog Circuit for Timing Measurements With Large
Area SiPMs Coupled to LYSO Crystals''}, {\bf IEEE TRANSACTIONS ON NUCLEAR SCIENCE},  Volume: 60, Issue:2, Part:2 (2013)
\bibitem{Ambrosi} G.~Ambrosi et al., {\it ``INFN Camera demonstrator for the Cherenkov Telescope Array''}, these proceedings

\end{thebibliography}
\end{document}